\documentclass[]{aa}  
\usepackage{graphicx}
\usepackage{txfonts}
\usepackage{subfigure}

\usepackage{natbib}
\bibpunct{(}{)}{;}{a}{}{,}

\begin{document}
   \title{Thin shell morphology in the circumstellar medium of massive binaries}
\titlerunning{Thin shell morphology around massive binaries}

   \author{A. J. van Marle \inst{1}
         \and
          R. Keppens\inst{1,2,3}
        \and
          Z. Meliani\inst{1}
          }

   \institute{Centre for Plasma Astrophysics, K.U. Leuven,
              Celestijnenlaan 200B, B-3001 Heverlee, Belgium\\
              \email{AllardJan.vanMarle@wis.kuleuven.be} 
         \and
             FOM Institute for Plasma Physics Rijnhuizen, P.O. Box 1207
           NL-3430 BE  Nieuwegein, the Netherlands
        \and
            Astronomical Institute, Utrecht University, Budapestlaan 6
            NL-3584 CD, Utrecht, the Netherlands
             }

   \date{??; ??}

 
  \abstract
   {In massive binaries, the powerful stellar winds of the two stars collide, leading to the formation of shock-dominated environments that can be modelled only in 3D.}
   {We investigate the morphology of the collision front between the stellar winds of binary components in two long-period binary systems, one consisting of a hydrogen rich Wolf-Rayet star (WNL) and an O-star and the other of a Luminous Blue Variable (LBV) and an O-star. 
Specifically, we follow the development and evolution of instabilities that form in such a shell, if it is sufficiently compressed, due to both the wind interaction and the orbital motion.}
   {We use MPI-AMRVAC to time-integrate the equations of hydrodynamics, 
   combined with optically thin radiative cooling, on an adaptive mesh 3D grid. 
  Using parameters for generic  binary systems, we simulate the interaction between the winds of the two stars.}
   {The WNL + O star binary shows a typical example of an adiabatic wind collision. 
The resulting shell is thick and smooth, showing no instabilities. 
On the other hand, the shell created by the collision of the O star wind with the LBV wind, combined with the orbital motion of the binary components, is susceptible to thin shell instabilities, which create a highly structured morphology. 
We identify the nature of the instabilities as both linear and non-linear thin-shell instabilities, with distinct differences between the leading and the trailing parts of the collision front. We also find that for binaries containing a star with a (relatively) slow wind, the global shape of the shell is determined more by the slow wind velocity and the orbital motion of the binary, than the ram pressure balance between the two winds. }
   {The interaction between massive binary winds needs further parametric exploration, to identify the role and dynamical importance of multiple instabilities at the collision front, as shown here for an LBV + O star system.}


   \maketitle
%

\section{Modelling circumstellar environments}
\label{sec-intro}
Massive stars lose a significant amount of their mass during their evolution. 
The mass is lost in the form of both quasi-continuous winds and sporadic eruptions. 
Changes in mass-loss rate and velocity create a series of hydrodynamical interactions, shaping the circumstellar medium. 
The result of these interactions can be observed in the form of circumstellar nebulae, which are temporary high density structures surrounding the star. 

The formation of such structures around single stars has been successfully modelled in quite some detail \citep[e.g.][]{GarciaSeguraetal:1996a,GarciaSeguraetal:1996b,Freyeretal:2003,Freyeretal:2006,vanMarleetal:2005,vanMarleetal:2007,vanMarleetal:2008}. 
However, since most massive stars occur in binaries, the value of such single star simulations is limited. 
At the same time, single star circumstellar bubbles can be studied in restricted dimensionality, allowing very high resolution parametric
studies, highlighting the role of various thin-shell instabilities in circumstellar morphology. 
For a binary, even the symmetric case of two identical stars requires 3D computations, at considerably
increased computational costs.

Recent work exploiting Smooth Particle Hydrodynamics (SPH) methodologies \citep[e.g.][]{Okazakietal:2008} investigated the 3D wind-wind interactions believed to occur in the \object{Eta Carinae} star system. 
However, standard SPH algorithms face difficulties in the vicinity of strong shocks, while the collision between two strong stellar winds inherently leads to their creation. 
Modelling the wind collision using grid-based methods has usually focused on the details of the interaction front in between the two stars, such as in the work by \citet{Walder:1998,  FoliniWalder:2000}. 
More recently, models emerged where both symmetric and asymmetric systems of massive O-stars, on circular or elliptic orbits, were followed in detail \citep{Pittard:2009,PittardParkin:2010}. \citet{Pittard:2009} in particular uses Riemann-solver based discretizations, on a fixed 3D grid, solving hydrodynamics together with optically thin radiative losses (along with gravity and radiative driving terms), and directly relates to the work presented here. 
While those simulations have already shown the complex morphology of the circumstellar medium, we augment those results here for a hydrogen rich Wolf-Rayet (WNL)+O binary and a Luminous Blue Variable (LBV)+O binary, paying specific attention to the role and nature of the occurring thin-shell instabilities.

In this paper we focus on the circumstellar medium of  massive binaries with a relatively wide orbit taking the orbital period equal to 1 year. 
For the components of the binary system we choose a combination of a WNL and an O-type star and a combination of a LBV and an O-star, using generic parameters rather than attempting to simulate existing binary systems. 
To facilitate a comparison between the two systems, we use identical stellar mass and orbital parameters for both. 
This can be interpreted as two stages in the evolution of a massive binary, with the most massive of the two stars making the transition from hydrogen rich Wolf-Rayet to Luminous Blue Variable \citep{Langeretal:1994}.

In the region where the winds collide, a shell forms, which is strongly compressed by the collision. 
For the WNL + O star binary the collision is adiabatic on both sides resulting in a thick, smooth shell. 
However, for the second binary, owing to the disparate nature of the winds (the LBV wind is relatively slow and quite dense, whereas the O-star wind is ten times faster, but has comparatively low density), the nature of the shock is radiative on one side and at least partially adiabatic on the other. 
Since the resulting shell is thin, as a result of efficient radiative cooling, it will be subject to combined hydro and radiative instabilities, depending on the (evolving) wind parameters. We identify the nature of the instabilities in the specific LBV+O scenario, which form in the shell due to both the wind interaction and the stellar orbital motion. 
In particular, the shell is characterized by the presence of multiple thin-shell instabilities, with clear differences between leading and trailing interaction fronts.

  \begin{table*}
      \caption[]{Binary parameters. All parameters regarding the primary star denoted with subscript 1, all parameters regarding secondary with subscript 2.}
         \label{tab:bin}
\centering
\begin{tabular}{ccccccccc}
\hline\hline
Simulation & Mass$_1$       & Mass$_2$    & $\dot{M}_1$       & $\dot{M}_2$       & $\varv_1$ & $\varv_2$ & Period & eccentricity   \\
           & [M$_\odot$]    & [M$_\odot$] & [M$_\odot /yr]$   & [M$_\odot /yr]$   & [km/s]    & [km/s]    & [yr]   &                \\
\hline
\\
WNL+O      &  50            & 30          & $5\times 10^{-6}$ & $5\times 10^{-7}$ &  1500     & 2000      & 1      & 0               \\
LBV+O      &  50            & 30          & $1\times 10^{-4}$ & $5\times 10^{-7}$ &   200     & 2000      & 1      & 0               \\
\hline
\end{tabular}
 \end{table*}

\section{Generic binary models}
\label{sec-method}
\subsection{Governing equations}
We simulate the hydrodynamical interaction between the winds of two massive stars, solving for the dynamics in the center of mass frame of the two stars. 
The stellar masses, orbit and wind parameters are shown in table~\ref{tab:bin}. 
The O-star and WNL wind mass-loss rate are chosen rather low, following the findings of \citet{Bouretetal:2005,Vinkdekoter:2005,Mokiemetal:2007} and others that clumping has led to overestimation of the mass-loss rate of hot stars. 
The LBV mass-loss rate reflects values found by \citet{Vinkdekoter:2002}. 

Since we simulate a binary with a relatively wide orbit, we can ignore the acceleration zone of the stellar winds 
\citep[typically$<10$R$_\star$][and citations therein]{Lamerscassinelli:1999} and model wind-wind interactions between winds at terminal velocity. 
This also justifies the neglect of gravity and radiation-driven body forces, which were taken along in the work by \citet{Pittard:2009}. 
The changing orbital positions are calculated following Kepler's law and the terminal wind zones are prescribed as explained below. 
We take into account the effect of optically thin radiative cooling, using the cooling curve for solar metallicity from \citet{MellemaLundqvist:2002}. 
Since the medium close to two massive stars can be assumed to be photo-ionized, we keep the gas at a minimum temperature of 10\,000~K throughout the entire computational domain.

We solve the usual equations of hydrodynamics: conservation of mass, 
\begin{equation}
\frac{\partial \rho}{\partial t} ~+~\nabla \cdot (\rho {\mathbf v} ) ~~=~ 0, 
\label{eq:mass}
\end{equation}
with $\rho$ the mass density and $\mathbf{v}$ the velocity vector; conservation of momentum, 
\begin{equation}
\frac{\partial \rho\mathbf{v}}{\partial t} ~+~\nabla \cdot (\rho {\mathbf v}{\mathbf v} ) ~~=~ -\nabla p, 
\label{eq:momentum}
\end{equation}
with $p$ the thermal pressure, and with the energy equation given by
\begin{equation}
\frac{\partial e}{\partial t} ~+~\nabla \cdot (e {\mathbf v} ) ~+~ \nabla \cdot (p{\mathbf v}) ~=~ -n^2 \Lambda(T), 
\label{eq:energy}
\end{equation}
with $e$ the total energy density, and energy loss due to radiation in the righthand side. 
This term depends quadratically on the ion density $n$, and $\Lambda(T)$ is the temperature-dependent cooling rate for solar metallicity from \citet{MellemaLundqvist:2002}, 
which is similar to the SPEX-code based cooling curve used by \citet{Pittard:2009}, but incorporates fewer ion species.

\subsection{Numerical setup}
We use the MPI-AMRVAC code \citep{Melianietal:2007}, using 
its hydrodynamical module, on grids where the local resolution can be changed through 
adaptive mesh refinement (AMR). 
We implemented optically thin radiative cooling using the new exact integration method from \citet{Townsend:2009}, which improves calculation speed and numerical stability. A comparison by \citet{Vanmarlekeppens:2010} between different source term treatments for these optically thin radiative loss terms, in 1D to 2D single wind bubble evolutions, clearly showed the need for AMR combined with robust radiative cooling prescriptions. 
For the hydro, we use a shock-capturing TVDLF scheme \citep{TothOdstrcil:1996}, employing limited linear reconstructions.

Our grid consists of a flat slab measuring $2.5\times10^{14}$~cm along the X- and Y-axes (the orbital plane of the binary) with 240 grid points along each axis in the crudest refinement level and $2.5\times10^{13}$~cm along the Z-axis with 20 grid points at its lowest resolution.  
For this particular simulation this means that in the X-Y plane the grid covers approximately 4 times the orbital separation along each axis. 
Assuming a typical O-star radius of about 10\,R$_\odot$, this amounts to 350\,R$_{\mathrm{O-star}}$. 
The radius of WNL and LBV stars is less well-known, as these stars have optically thick winds. 
Therefore, the effective temperature is not defined at the stellar surface, but somewhere in the wind region. 
Typically though, the WNL radius will be somewhat larger than the radius of an O-star while the LBV radius will be about an order of magnitude larger than R$_{\mathrm{O-star}}$.

We then use two additional grid levels, where the intermediate level achieves
an effective resolution of $480\times 480\times40$, which is controlled by the automated error estimation procedure based on density variation, while the top
grid level
is at all times enforced in the direct vicinity of the two stars, making the
effective resolution there $960\times 960\times80$. The latter grid refinement level is thus activated in a fully user-controlled manner, employing the knowledge of the instantaneous stellar positions.

At the beginning of each time step, this orbital position is calculated from the orbital parameters, and a sphere around each star of $5\times10^{12}$~cm in diameter is filled with free-streaming wind material. This approach mimics the work by~\citet{Pittard:2009}, while external to this radius, the wind expands freely according to the governing hydrodynamics.

   \begin{figure}
   \centering
   \includegraphics[width=\columnwidth]{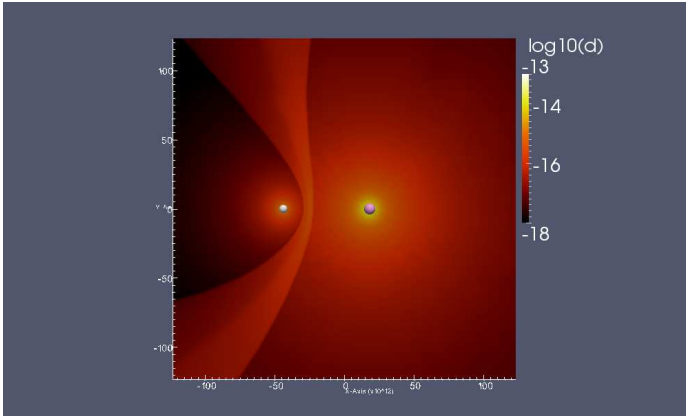}
      \caption{Gas density in g/cm$^3$ of the WNL+O binary system after one full orbital revolution (1\,yr). Owing to the higher ram pressure of the WNL (violet sphere) wind the bowshock curves around the O-star (white sphere).  
The shell is thick on both sides of the contact discontinuity and shows no sign of instabilities. 
              }
         \label{fig:WRdens}
   \end{figure}

   \begin{figure}
   \centering
   \includegraphics[width=\columnwidth]{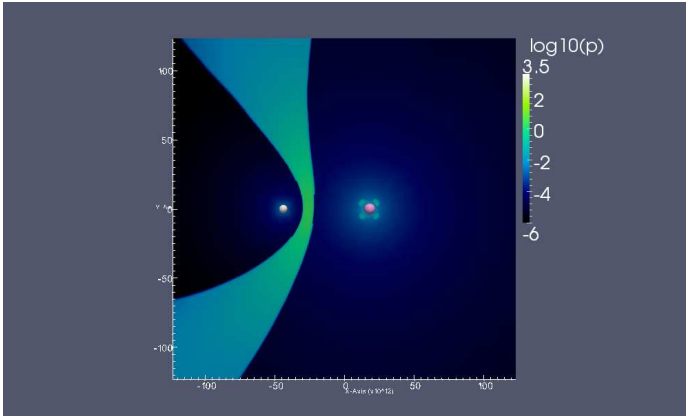}
      \caption{Thermal pressure in erg/cm$^3$ for the WNL+O binary system at the same time as Fig.~\ref{fig:WRdens}. 
The thermal pressure of the shocked wind region balances against the ram pressure of the two winds. 
              }
         \label{fig:WRpres}
   \end{figure}

\section{Circumbinary medium morphology}
\label{sec-result}
\subsection{WNL+O binary}
The morphology of the circumbinary medium of the WNL+O binary, as shown in Figs.~\ref{fig:WRdens} and \ref{fig:WRpres} is a typical example of a purely adiabatic wind-wind collision \citep{Stevensetal:1992}. 
The shock is adiabatic on both sides of the contact discontinuity, resulting in a thick shell comprised of shocked wind gas. 
The edges of the shell are defined by the point where the thermal pressure inside the shocked wind region is equal to the ram pressure of the wind. 
The WNL wind, which has a much higher ram pressure (a factor of 7.5) clearly dominates the collision and has folded the O-star wind back into a bow-shock.  
Owing to its thickness, the shell is not subject to thin-shell instabilities and remains smooth. 

The leading edge of the shell (lower part in Figs.~\ref{fig:WRdens} and \ref{fig:WRpres}) runs approximately parallel to the WNL wind, whereas the trailing edge falls slightly behind due to the orbital motion, which runs counter-clockwise.  
Since there is a shear along the edges of the shell it would, in theory, be subject to Kelvin-Helmholtz instabilities. 
However, the probability of such instabilities occurring is reduced by the orbital motion of the shell, which is perpendicular to its surface and tends to wipe out such instabilities.

   \begin{figure}
   \centering
   \includegraphics[width=\columnwidth]{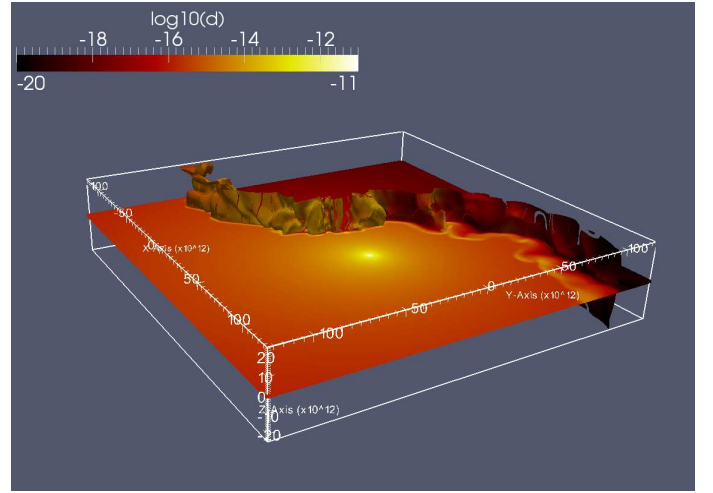}
      \caption{3D image of the circumstellar medium around an LBV + O binary, after one orbital period (1\, yr). 
      The slice along the equatorial plane shows the logarithm of the density. 
      An isosurface along which the absolute velocity equals 300\,km/s is representative of conditions along the contact discontinuity between the shocked LBV wind and the shocked O-star wind. 
      N.B. the z-axis has been stretched by a factor 2 to show more clearly the 3D nature of the instabilities.
}
         \label{fig:binary_3D}
   \end{figure}

   \begin{figure}
   \centering
   \includegraphics[width=\columnwidth]{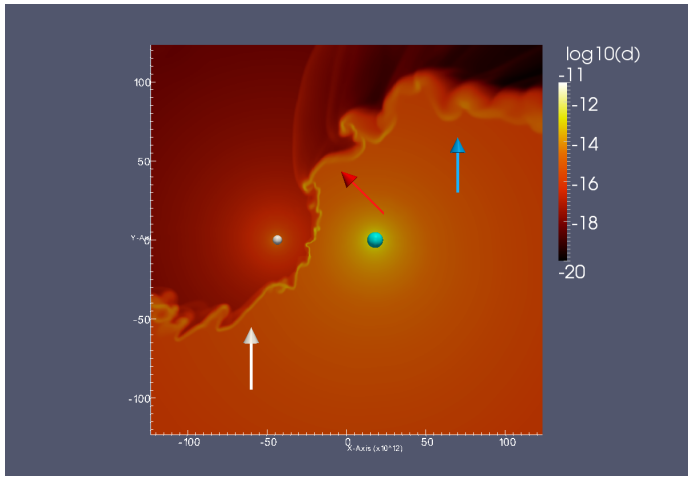}
      \caption{Density of the circumstellar medium in the LBV + O system at the same moment in time as Fig.~\ref{fig:binary_3D}.
      This figure shows a cross-section in the orbital plane of the binary. 
      The wind from the Luminous Blue Variable (blue sphere) dominates due to its higher ram-pressure. 
      The shell created by the collision is unstable, showing two kinds of thin shell instabilities:
      non-linear order thin-shell,  instabilities at the head of the bow shock around the O-star (white sphere) and in the region directly between the two stars. 
      Downstream (red arrow) the instabilities are of linear thin-shell type. 
      Even further downstream, after the shell has made its curve around the approaching LBV star the shell begins to diffuse due to its own internal pressure.
              }
         \label{fig:dens}
   \end{figure}
   \begin{figure}
   \centering
   \includegraphics[width=\columnwidth]{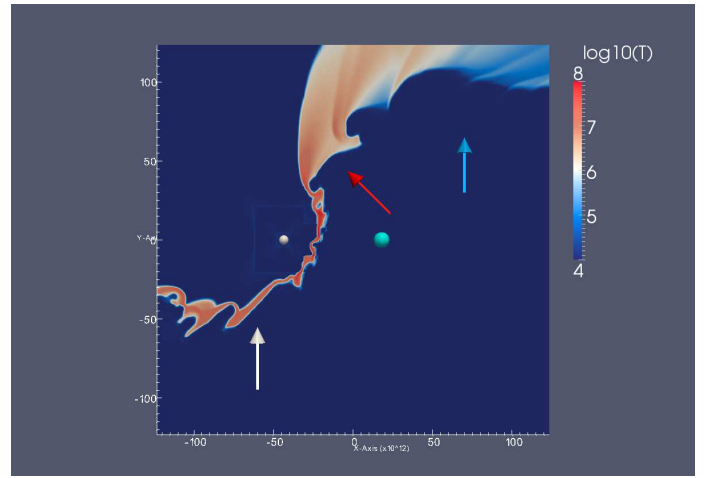}
      \caption{Temperature of the circumstellar medium  in K at the time from Figs.~\ref{fig:binary_3D} and \ref{fig:dens}. 
      This figure shows the reason for the split in thin-shell instabilities. 
      At the front of the bowshock as well as between the two stars, the thermalized zone is extremely thin, as a result of effective radiative cooling. 
      The shock, owing to its high density and temperature is almost completely radiative there, making the shell subject to ram-pressure on both sides. 
      In the trailing end (red arrow) the shock interaction is less violent and densities are lower. As a result the shock is no longer radiative and a relatively thick thermalized zone forms. 
      Hence the zone between contact and termination shock for the LBV wind bubble feels ram-pressure on one side and thermal pressure on the other. 
      This changes the nature of the thin-shell instabilities. 
              }
         \label{fig:temp}
   \end{figure}

   \begin{figure}
   \centering
   \includegraphics[width=\columnwidth]{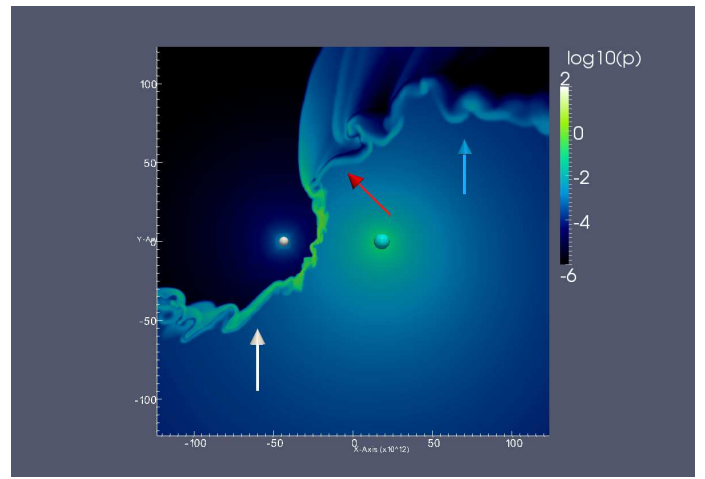}
      \caption{Thermal pressure in erg/cm$^3$ of the circumstellar medium at the same  time as Figs.~\ref{fig:binary_3D} through \ref{fig:temp}. 
This plot shows how the thermal pressure supporting the shell from the O-star side disappears at a larger distance (blue arrow). 
The shell still feels the ram pressure from the LBV star wind, which pushes it ahead, but without a counter force from the other side 
it is no longer compressed and starts to expand under its own internal pressure. 
This plot also shows how the O-star wind forms bowshocks around the individual deformations in the downstream region.
              }
         \label{fig:pres}
   \end{figure}

\subsection{LBV+O binary}
The morphology of the circumstellar medium around the LBV+O binary system is presented in Figs.~\ref{fig:binary_3D} through \ref{fig:pres} and clearly shows a much more complicated structure due to instabilities in the shell. 
Figure~\ref{fig:binary_3D} gives an impression of the 3D structure through a constant absolute velocity isosurface, at a value that selects the contact discontinuity between the two shocked winds. 
Since this follows the contour of the shell, it shows the intrinsic three-dimensional structure of the instabilities. 
Figures~\ref{fig:dens} through \ref{fig:pres} show 2D slices in the orbital plane, taken after a full orbital evolution. 
The LBV wind, which has the higher ram-pressure (${P_{\mathrm{ram}}({\mathrm{LBV}})/P_{\mathrm{ram}}({\mathrm{O}})= 20}$), 
dominates the interaction, so a bow-shock forms ahead of the O-star as it plows into the LBV wind, 
but unlike the WNL+O binary where this bow-shock looks almost symmetric, the LBV+O binary shows that the trailing edge of the bow-shock has fallen behind in the orbital motion, despite the fact that the LBV wind has a higher ram pressure than the WNL star from the first simulation. 
This is the result of the low velocity of the LBV wind and will be discussed in more detail in Section~\ref{sec-shell}.

\begin{figure*}
 \centering
\mbox{
\subfigure
{\includegraphics[width=0.5\textwidth]{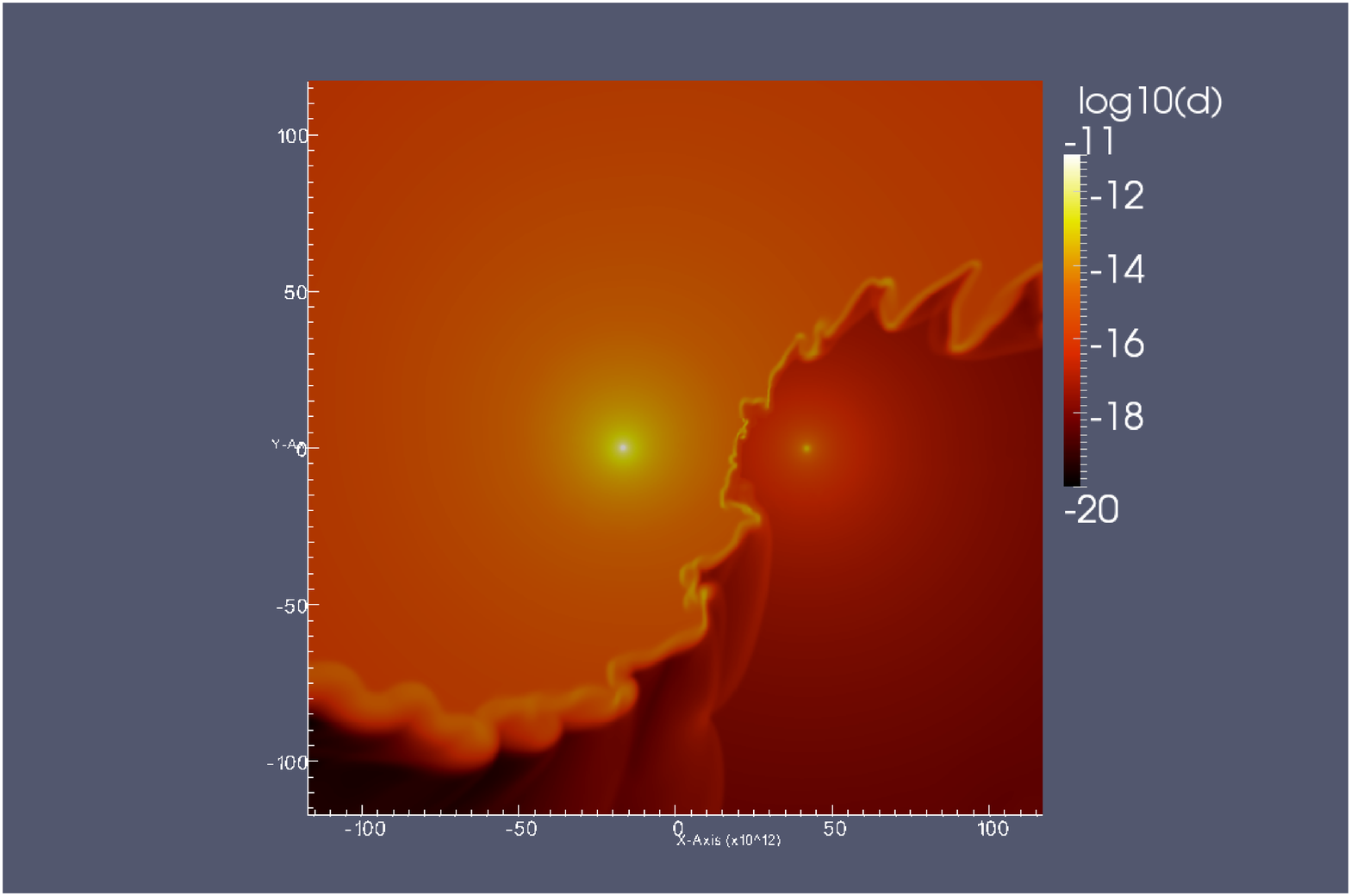}}
\subfigure
{\includegraphics[width=0.5\textwidth]{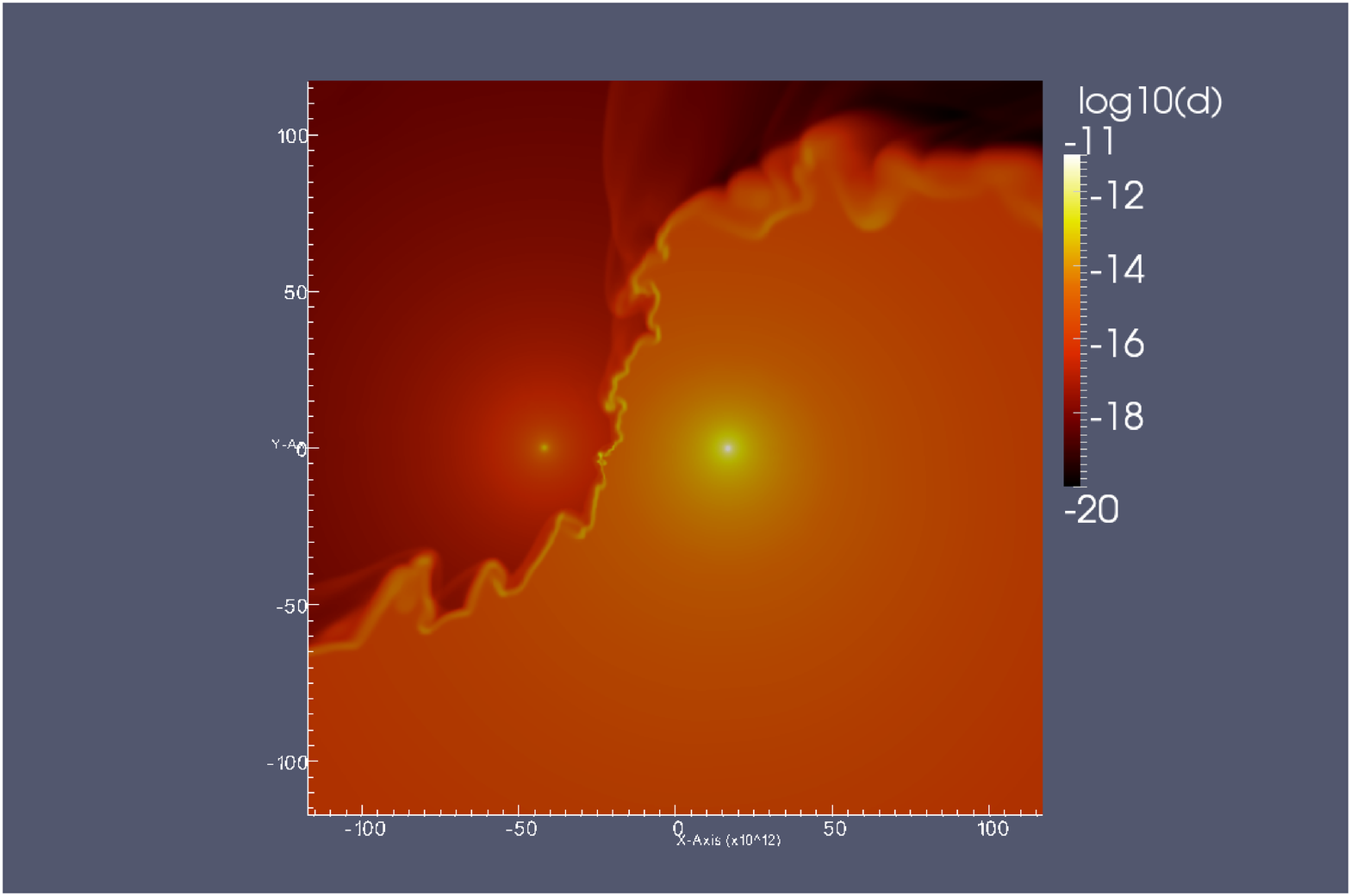}}}
\caption{Density in g/cm$^3$ of the circumbinary medium of the LBV + O binary after 1.5 (left) and 2 (right) years. The general structure of the shell does not change over time, though the individual deformations vary.}
 \label{fig:dens23}
\end{figure*}

Figure~\ref{fig:dens23}  demonstrates the evolution of the circumstellar medium over time as the LBV+O binary completes its second orbit, showing the density of the gas after 1.5 (left) and 2 (right) orbits respectively. 
The general morphology of the shell remains the same, though the thin-shell instabilities at the front of the bow-shock become a bit more pronounced. 
This is further demonstrated in Fig.~\ref{fig:shelltime}, which shows the development of the LBV+O binary shell over time. 
It presents isosurfaces defined at an absolute velocity of 300\,km/s (see also Fig.~\ref{fig:binary_3D}) at 5 subsequent moments in time, covering one complete orbit. 
The first contour (red) shows the shell 1/5 of an orbit after Figs.~\ref{fig:binary_3D} through \ref{fig:pres} and the final contour (white) shows the shell at the same time as Fig.~\ref{fig:dens23}. 
The global shape of the shell remains constant over time, rotates rigidly around the center of mass of the binary. 
The local instabilities vary over time, but maintain the same general shape, with relatively small deformations in the region directly between the two stars and much larger in the leading and trailing parts. 
At the leading end of the shell the deformations also maintain their `saw-tooth' appearance.
The distortions of the shell directly between the two stars remain comparatively small. This is due to two effects: 
As the distortion moves toward either of the stars it will experience an increasing ram pressure owing to the increase in wind density, which slows down its motion. 
Also, the growth time of these instabilities is on the order of the orbital period, so the shell itself moves significantly before the instability has time to grow further. 
The nature of these instabilities is explored in more detail in Section~\ref{sec-discus}.

\section{The collision region}
\subsection{Nature of the shock}
\label{sec-shock}
As shown by \citet{Stevensetal:1992}, the structure of the interaction region is determined by the nature of the shocks. 
For an adiabatic shock an extended shocked wind region with a high 
temperature will develop between the free-streaming wind and 
the contact discontinuity. 
On the other hand, if the shock is radiative, the shocked wind region will be compressed into a thin 
shell. 
The defining criterion for colliding wind shocks is the cooling parameter, which \citet{Stevensetal:1992} determined as 
the relation between the cooling timescale $t_{\mathrm{cool}}=kT_s/(n_w \Lambda(T_s))$  and the escape time from the shocked region $t_{\mathrm{esc}}=d/c_s$, 
with $k$ the Boltzmann constant, $T_s$ the shocked wind temperature, $n_w$ the particle density in the wind at the shock, $d$ the distance to the star and $c_s$ the post-shock sound speed. 
Assuming that $c_s\sim\varv\sim\sqrt{T_s}$ and taking $\Lambda$ constant for typical shock temperatures of $10^7...10^8\,$K, this gives us a cooling parameter of: 
\begin{equation}
\chi~=~\frac{t_{\mathrm{cool}}}{t_{\mathrm{esc}}}~\approx~\frac{\varv_8^4 d_{12}}{\dot{M}_{-7}},
\end{equation}
with $\varv_8$ the wind velocity in $10^8$\,cm/s, $d_{12}$ the distance between star and shock in $10^{12}$\,cm and $\dot{M}_{-7}$ the wind mass-loss rate in $10^{-7}$\,M$_\odot$/yr. 
Although this equation is not very precise (it neglects the temperature dependence of the radiative cooling), it gives an indication as to the nature of the shock. 

For the WNL+O binary, the cooling parameter for the O-star is: $\chi\,\sim\,50$ in the region directly between the two stars, whereas for the WNL star it is $\chi\,\sim\,4$, indicating that both shocks are at least partially adiabatic, as $\chi>1$.
This is confirmed by the compression {\bf factor}, which for both shocks is approximately 4.
Obviously, further away from the collision point the shocks becomes even more adiabatic as the distance to the star increases leading to lower densities and less efficient cooling.

For the second binary, directly between the two stars the LBV wind has ${\chi\,\sim\,10^{-4}\ll\,1}$, which makes this shock completely radiative, resulting in a thin shell of stalled LBV wind material. 
The O-star wind again has ${\chi\,\sim\,50>\,1}$, which makes the shock adiabatic. 
Compression rates vary due to the local instabilities. 
However, though the O-star wind does develop a thermalized layer between the free-streaming wind and the contact discontinuity, this layer is comparatively thin (approx. twice the cross-section of the compressed LBV wind). 
This is of no great consequence for the WNL+O binary, which doesn't show instabilities, but is crucial in the LBV+O binary.

This calculation does not take into account the orbital motion of the stars, which greatly complicates the situation  for the LBV+O binary, where the slower of the two winds (the LBV wind) moves with a speed that is comparable to the orbital velocity of the rigidly rotating bowshock. 
For the WNL+O binary the shell is effectively caught between the ram pressure of both winds along its entire length. 
However, for the LBV +O binary the situation is completely different. 
At the edge of the bow-shock, where it hits the LBV wind material (white arrow in Figs.~\ref{fig:dens} through \ref{fig:pres}), 
the shell is bent further toward the O-star due to the higher ram pressure of the LBV wind as well as the fact that it is sweeping up material that is much denser than in the case of the WNL wind. 
As a result, the relevant collision velocity on the LBV wind side of the contact discontinuity is not the LBV wind speed, which runs almost parallel to the shock front, but the orbital velocity of the shock into the wind material. 
At the trailing end of the bow shock (red and blue arrows in Figs.~\ref{fig:dens} through \ref{fig:pres}), the shell has to keep up with the orbital rotation while it is being driven ahead by the slower of the two winds (either the WNL or LBV wind). 
This can result in the loss of the shock as will be discussed in Section~\ref{sec-shell}.

   \begin{figure}
   \centering
   \includegraphics[width=\columnwidth]{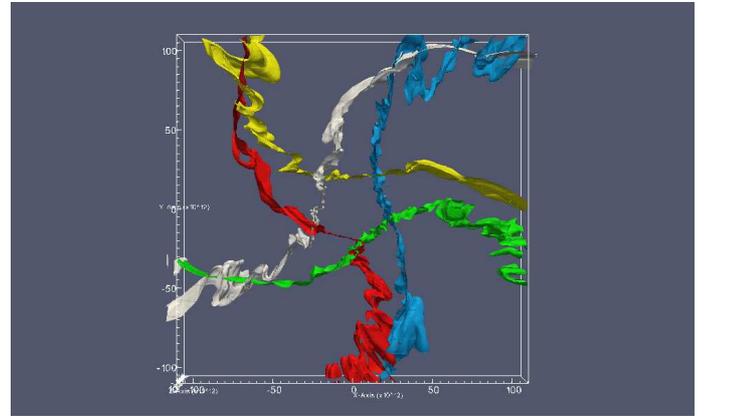}
      \caption{The shell between the LBV and O star at five different points in time, starting 1/5$^{\rm th}$ orbit after Figs.~\ref{fig:binary_3D} through \ref{fig:pres} (red) and at each 1/5$^{\rm th}$ orbit afterwards (green, blue, yellow) The final shell (white) coincides with the righthand plot in Fig.~\ref{fig:dens23}. 
The shells are defined as an isosurface contour drawn at a velocity of 300\,km/s (see also Fig.~\ref{fig:binary_3D}). 
The global shape of the shell remains constant over time, rotating rigidly around the center of mass of the binary (0,0). 
Local instabilities vary, but retain the same general characteristics, being small between the stars, but much larger in the advancing and trailing ends of the shell.
}
         \label{fig:shelltime}
   \end{figure}

\subsection{The loss of ram pressure balance in the trailing part of the shell}
\label{sec-shell}
The shape of the trailing end of the collision region depends on the relative velocities of orbital motion and the stronger of the two winds.
This wind drives the trailing end of the bow shock ahead, which means that the shell can never move faster than the component of the LBV wind velocity perpendicular to the shell (or even 1/4$^{\rm th}$ of this velocity as long as a shock is maintained  at the collision front, because of the Rankine-Hugoniot conditions). 
Since the shell is, effectively, rigidly rotating around the center of mass of the binary (see also Fig.~\ref{fig:shelltime}), its orbital velocity has to increase linearly with the distance to the center of mass ($\varv_{\mathrm orbit}=2\pi R\Omega$). 
Therefore, for larger radius the orbital velocity approaches the LBV wind velocity. 
At this point the shell has to fall back to catch a larger percentage of the LBV wind momentum (which would otherwise hit it at a very shallow angle). Eventually even this fails and the shell starts to lag behind the orbital motion.  
A similar effect can be observed in some of the simulations by \citet{Pittard:2009}.
The WNL wind, which is 7.5 times faster than the LBV wind, only reaches this point at the edge of our computational domain, which can be seen in the gradual change in the curvature of the shell in Figs.~\ref{fig:WRdens} and \ref{fig:WRpres} at a distance of about $10^{14}$cm from the center of mass. 
This agrees with the description by \citet{Parkinpittard:2008}, which shows that the angle of the bow-shock with the axis between the two stars depends on the velocity of the slowest of the stellar winds, relative to the orbital motion.

\begin{figure}
   \centering
   \includegraphics[width=\columnwidth]{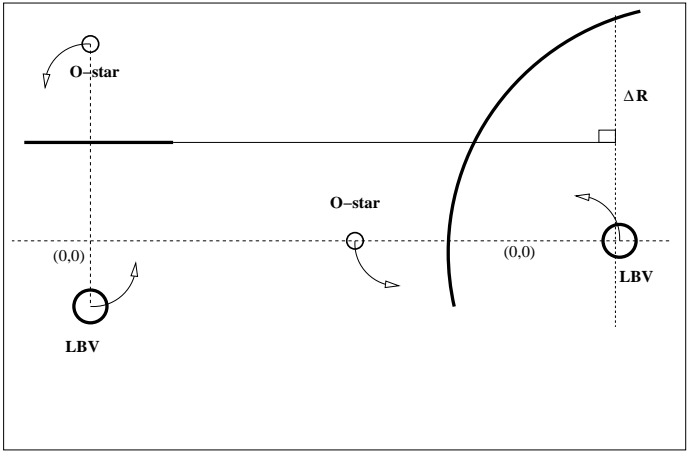}
      \caption{Displacement of the shell over a 1/4$^{\rm th}$ orbit (0.25\,yr). The distance $\Delta\,R$ over which the shell moves is determined by its own inertia, the remnant of thermal pressure that balances it against the ram pressure of the approaching LBV star and the velocity of the LBV wind, placing an upper limit on the shell velocity.
              }
         \label{fig:binary_sketch_2}
   \end{figure}

The WNL+O binary maintains ram pressure balance between the two winds in the entire computational domain. 
For the LBV+O binary the situation is completely different. 
Downstream from the collision (red arrow in Figs.~\ref{fig:dens} through \ref{fig:pres}), the shocked wind region effectively no longer feels the ram pressure of the O-star wind (which runs parallel to the shock), allowing the shocked wind region to expand.  
We can set limits for the shell displacement. 
If we assume that enough thermal pressure always remains to counterbalance the ram pressure of the LBV wind, then the shell stays at a constant distance to the LBV star and its displacement equals the motion of the LBV star perpendicular to the shell. 
This is the absolute minimum displacement. 
The maximum can be deduced from a hypothetical case in which there is no thermal pressure at all and the shell is simply carried along by the LBV wind. 
In that case the displacement is equal to the LBV wind velocity times the time interval. 
Obviously this overestimates the motion, since the mass of the shell itself (and its related inertia) would slow it down even without any thermal pressure. 
Using these two estimates we obtain for a quarter orbit (90$^o$, see Fig.~\ref{fig:binary_sketch_2}),
\begin{eqnarray}
\Delta\,R_{\mathrm{min}}~&=&~R_{\mathrm{orbit}}(\mathrm{LBV})~=~1.75\times\,10^{13}\,\mathrm{cm}, \\
\Delta\,R_{\mathrm{max}}~&=&~\varv(\mathrm{LBV})\,\Delta\,t~=~1.58\times\,10^{14}\,\mathrm{cm}.
\end{eqnarray}
Clearly, these two limits lie far apart and the actual displacement of the shell in our simulation, which is about $5\times\,10^{13}$\,cm relative to the center of mass of the binary (point (0,0) in Fig.~\ref{fig:binary_sketch_2}), falls in between. 
The Rankine-Hugoniot conditions determine that the post shock speed is 1/4$^{\mathrm{th}}$ of the pre-shock speed. 
Therefore, the shell could only move 1/4$^{\mathrm{th}}$ of $\Delta\,R_{\mathrm{max}}$ if a shock was maintained between the shell and the LBV wind. 
This would limit the displacement of the shell to $\Delta\,R=3.95\times\,10^{13}\,\mathrm{cm}$. 
Since the actual displacement is larger, we can conclude that the shock between the LBV wind and the shell has been lost. 


The point where the shell will deviate from the balance between the ram pressure of the two winds can be approximated by using the condition that 
the component of the LBV wind along the orbital direction of motion has to exceed the orbital velocity determined by the binary stars:
$\varv_\angle\,\geq\,2\pi\,R\Omega$, with $\varv_\angle$ the component of the LBV wind along the direction of orbital motion.
This condition has been worked out in Appendix~\ref{sec-turningpoint}. 
For our LBV+O binary the resulting equation is easy to use, since the shell between the two stars runs almost perpendicular to the axis connecting the two stars. 
Using an approximation where $\Delta\,X$ in Eq.~\ref{eq:v_condition} is constant (set to $2\times\,10^{13}$, the distance from the center of mass to the point where the two winds collide head-on), the turning point of the shell lies at $\Delta\,Y\simeq\,2.35\times10^{13}$cm. 
This coincides quite well with the point where our simulation shows a significant deviation from the bowshock in Figs.~\ref{fig:dens} through \ref{fig:dens23}. 
From this point on the shell starts to curve backwards into the LBV wind region until, eventually, it runs parallel to the orbital motion.

This calculation does not take into account the distortion of the shell due to instabilities. Even a comparatively small instability can effectively shield a large section of the shell from the ram pressure of the O-star wind as a bow-shock forms around the instability as can be seen in Figs.~\ref{fig:dens} through \ref{fig:dens23}. 

As the shell curves away from the O-star wind (blue arrow in Figs.~\ref{fig:dens} through \ref{fig:pres}), because it cannot keep up with the orbital rotation, it loses the support from the thermalized O-star wind region and starts to expand under its own internal pressure. Since the O-star wind moves faster than the shell, the area behind the shell is essentially empty making it easy for the shell to expand.  
It gets pushed ahead by the wind of the approaching LBV star and accelerates away as the ram pressure of the wind overcomes the inertia of the shell.
For the WNL+O binary, this would only happen at a much larger distance from the binary stars, as the faster WNL wind is capable of maintaining a shock at higher orbital velocity.

\subsection{Ionization state}
Our assumption of complete photo-ionization is supported by the results. 
The particle density in the shell reaches a maximum of about $10^{10}$\,cm$^{-3}$. 
For a typical O-star, which produces on the order of $10^{49}$\,s$^{-1}$ high energy photons, this would imply a Str{\"o}mgren radius of about $5\times10^{13}$\,cm \citep[Eq.\,5.14 from][]{Dysonwilliams:1997} if the density was constant from the O-star to the outer edge of the shell. 
In reality the density around the O-star is much lower due to the free expansion region of the wind, leaving only the thin shell to absorb the photons since recombination in the free-streaming wind region is negligible \citep{Garciasegurafranco:1996}. 
This means that the O-star alone can ionize the shell between the two stars, even without the presence of the LBV star. 
Only at larger distances from the star (once the matter has been carried away downstream) will it recombine. 
Obviously, this is a generalization. 
Inside the shell high density clumps may form, which will be able to recombine due to self-shielding if they become optically thick. However, such structures will be small and can not be resolved by our current simulation. 
Moreover, this calculation fails to take into account the presence of dust which can shield the gas from the ionizing radiation.

\section{Multiple instability characterization for the LBV+O binary}
\label{sec-discus}
For the WNL+O binary both shocks are (near-)adiabatic, leading to a relatively thick shell that shows no instabilities. 
The LBV+O binary on the other hand reveals a complicated structure, as the highly compressed LBV wind forms a thin shell that is subject to thin-shell instabilities \citep{Vishniac:1983,Vishniac:1994}. 
That shells between colliding stellar winds are prone to these instabilities was found by \citet{Stevensetal:1992} in pioneering 2D simulations, which did not fully account for the effect of orbital motion. 
3D simulations by \citet{Dganietal:1996} showed similar results for the collision between a stellar wind and a parallel flow, as can occur if the star itself moves with supersonic speed relative to the interstellar medium. 

Our 3D simulation of two stars in orbit shows a more complicated picture, also found in some of the double O-star models analyzed by \citet{Pittard:2009}. 
For our LBV + O star system in circular orbit the shell instabilities come in two forms: 
non-linear thin-shell instabilities \citep{Vishniac:1994}, which occur if a thin shell is caught between ram-pressure from both sides, 
and linear thin-shell instabilities \citep{Vishniac:1983}, which depend on having ram-pressure on one side and thermal pressure on the other. 
As the winds from both stars are relatively cool ($\sim10\,000\,$K), the only way to create hot gas is if it gets thermalized by a shock. 
Whether such a thermalized zone forms, depends on the effectiveness of the radiative cooling, as discussed in Sect.~\ref{sec-result}.

Directly between the two stars the winds collide head on. 
The shock is completely radiative on the LBV side and at most marginally adiabatic on the O-star side. 
As a result the shocked LBV wind is compressed to a very thin shell (estimated compression {\bf factor} $\approx\,10...20$).
The shocked O-star wind is also compressed, but not to the same extent (estimated compression {\bf factor} $\approx\,5...10$). 
Initially, the instability is of the linear thin-shell variety as the shocked LBV wind shell is caught between the LBV wind ram-pressure and the thermal pressure from the shocked O-star wind. 
However, because of the long orbital period the instabilities have time to grow \citep[][eqn.~2.23 gives a growth rate of about 1\,month for this specific case, compared to the 1\,yr orbital period.]{Vishniac:1983}. 
As the instabilities cross the thin shocked O-star wind layer ($\sim5\times\,10^{12}$\,cm), the reverse shock in the O-star wind will start to conform to their shape. 
At this moment the nature of the instabilities becomes non-linear as they now effectively feel ram-pressure from both sides.
The growth rate of such instabilities is no longer determined by the sound speed in the shell, but by the velocity of the wind (200\,km/s for the LBV wind) which is faster than both the sound speed and the orbital motion of this part of the shell ($\leq\,50$\,km/s), allowing the instabilities to keep growing. 

\begin{figure}
   \centering
   \includegraphics[width=\columnwidth]{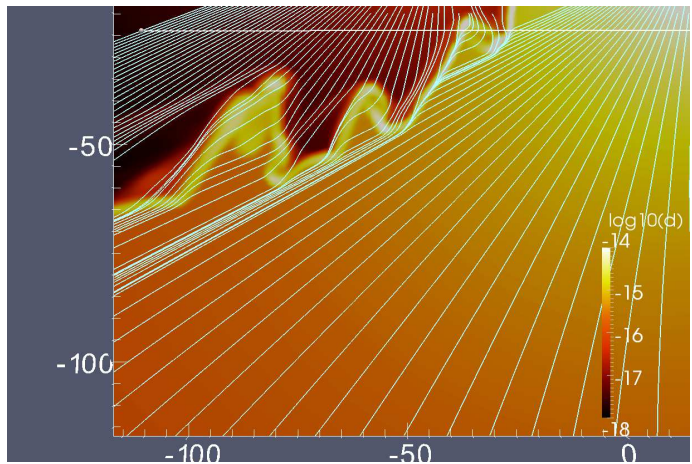}
      \caption{Streamtraces of the gas velocity in the orbital plane at the same time as in the right panel of Fig.~\ref{fig:dens23}. The instabilities deflect the LBV wind. There is no sign of Kelvin-Helmholtz instabilities (no circular motion). 
Instead, the asymmetric shape of the instabilities is caused by the shear of the LBV wind running parallel to the shell.
              }
         \label{fig:stream}
   \end{figure}

At the front of the bow shock (white arrow in Figs.~\ref{fig:dens} through \ref{fig:pres}) the shell does not actually feel the LBV wind, which moves nearly parallel to its surface. Instead, it feels a ram-pressure caused by its own movement (at an orbital velocity of about 90\,km/s) into the LBV wind region. 
However, as the development of the instability causes deformations that extend into the free-streaming LBV wind region, they are subjected to a side-ways force by the ram-pressure of the LBV wind, causing a strongly asymmetric shape, which reflects the relative forces acting on them. In the largest deformation, near the white arrow in Figs.~\ref{fig:dens} and \ref{fig:pres}, the slope facing the LBV wind is approximately four times as long as its displacement from the shell, conforming to the 4:1 ratio between the LBV wind ram-pressure and the orbital motion ram-pressure ($v_{\mathrm{LBV−wind}}\approx2\varv_{\mathrm{orbit}}$ and $P_{\mathrm{ram}}\sim\varv^2$.)
The other instabilities in this region are less asymmetric as the first instability shields them at least partially from the LBV wind. 
These instabilities look somewhat similar to Kelvin-Helmholtz instabilities, but a close-up of this same region at the same time as Fig.~\ref{fig:dens23}~(right) shows that there is no circular motion in them (Fig.~\ref{fig:stream}). 
This figure shows streamtraces of the velocity in the equatorial plane. 
The wind velocity clearly dominates but is deflected by the local instabilities. 
The interface between the shell and the LBV wind meets the criteria for Kelvin-Helmholtz instabilities due to the shear between the shell and the wind, but the shell moves perpendicular to the shear interface, effectively crushing such instabilities even if they start to form.
Owing to the bow-shocks that form around the instabilities on the O-star side of the shell, pockets of hot gas form behind the individual deformations, which therefore start to feel thermal pressure, rather than purely ram pressure.
This adds a linear thin-shell instability component. 
As a result the typical wavelengths here are longer than in the purely non-linear thin shell instabilities between the two stars \citep{Vishniac:1983,Vishniac:1994,GarciaSeguraetal:1996a}.

At the trailing end of the shock (red arrow in Figs.~\ref{fig:dens} through \ref{fig:pres}) the thermalized O-star wind forms an extensive layer, so that the shell only feels ram-pressure from the LBV wind, counterbalanced by thermal pressure from the shocked O-star wind. 
As a result, the instabilities here are becoming linear thin-shell instabilities, leading to longer wavelengths of the individual deformations of the shell. 
As these instabilities extend into the shocked O-star wind, they feel a shear force due to the velocity within the shocked O-star wind region, which runs parallel to the shell.  
This causes the beginning of Kelvin-Helmholtz type instabilities. 
In this region the temperature of the shell could drop below our own limit of 10\,000 K as the shocks here are relatively weak and local clumps may become dense enough for recombination. 
In that case it would be a good environment for dust formation as the local density is still high, but the temperature could drop to the order of a few thousand\,K, which would be acceptable for dust formation \citep{Cherchneffetal:2000}. 
Around the individual instabilities the O-star wind forms local bowshocks. 
This effectively decreases the thermal pressure on the shell, which starts to expand to the point where it can no longer be considered thin. It might be possible for individual clumps of dense matter from the shell to break off due to the shear force from the O-star wind and disperse into the low density region. 
However, this is difficult to predict from numerical simulations, since the resolution of the grid effectively determines the minimum size of a high density element.

Even further downwind (blue arrow in Figs.~\ref{fig:dens} through \ref{fig:pres}) the shell curves away from the O-star wind due to the orbital motion of the stars. 
At this point it no longer feels any sort of pressure from the O-star wind and therefore is no longer subject to first order thin-shell instabilities. 
The region behind the shell (where the O-star wind used to be) has a very low density, since the O-star wind is much faster than the motion of the shell and leaves a rarefied area into which the shell moves. 
It is being pushed ahead by the wind from the approaching LBV star, but without supporting pressure on the other side it is no longer being compressed and starts to diffuse as its own internal thermal pressure (primarily caused by the high density) causes it to expand. 
Since the high density parts have more inertia than the low density regions, they will tend to lag behind the low density regions, causing further distortion of the shell. 
At this point the transition between the LBV wind and the shell is no longer a shock, so the shell has effectively become a high density ridge, moving ahead of the wind.

\section{Conclusion and outlook}
\label{sec-concl}
Our simulation shows that the shell created by the collision of two different stellar winds can be subject to at least two kinds 
of thin-shell instabilities simultaneously, depending on the wind and orbital motion parameters. 
We find that the orbital motion plays a crucial part in shaping the interaction region between two stars, both on a global scale, in determining the position of the shell, and on a local scale, in determining the nature of the instabilities.

\subsection{Observational properties}
In terms of observations, the LBV+O binary would primarily radiate in the optical/IR. 
The (at least partially) adiabatic nature of the fast wind shock, will limit the production of X-rays. 
The radiatively cooling LBV wind shock would show a strong signature in the visual part of the spectrum (and possibly UV, though again, the high density of the LBV wind could absorb most of this). 
Due to the stability of the general shape of the interaction region, the optical signal, as produced by the shock, would probably be quite stable. Any strong variations would be more likely to be due to absorption by optically thick parts of the shell or LBV wind as they pass between the shock and the observer. 
The dense,  rapidly cooling shell can also enable dust formation, as observed in many massive binary systems. 
This leads to a clear IR signature as can be observed in \object{WR~140} \citep[e.g.][]{Monnieretal:2002, Varricattetal:2004} and the far more massive \object{Eta Carinae} \citep{Smith:2010}. 
The emission in optical will be increased due to the presence of extensive instabilities, which cause more kinetic energy to be transferred to thermal energy, which in turn radiates away through cooling. 
Since the instabilities enhance the local density in the shell, they will also increase the formation of dust, which in turn increases the IR radiation.

The WNL+O binary may produce more high energy photons as the fast WNL wind shock, though technically adiabatic, comes close to being radiative in the region directly between the two stars. 
Whether any such photons would be observable is another matter. 
They have to penetrate either the high density shell, or the free-streaming region of the WNL wind, which is at least partially optically thick. 
The result, if any X-ray emission would be seen at all, would depend very much on the angle of the observer with the shock cone and could lead to a periodic signal such as observed for \object{Eta Carinae} \citep{Corcoran:2005,Okazakietal:2008}, though in the case of \object{Eta Carinae} radiative braking of the wind may play an important role as well \citep{Parkinetal:2009} in determining the number of photons produced. 
This is not an issue in our binary where the winds only interact at terminal velocity.
The generally adiabatic nature of the shocks in the WNL+O binary prohibits the production of a strong optical signal and the luminosity of the massive stars themselves will tend to overwhelm any sign of the shock in the optical, unless the binary system could be spatially resolved. 
The most promising way to observe the shock would again be in the IR band, tracking the dust formed in the shell, though the amount of dust would be much less than for the LBV+O binary due to the lower densities in the shell and the fact that the WNL wind would contain less carbon than the LBV wind.

\subsection{Future work}
We here expand on the 2-D findings by \citet{Stevensetal:1992}, showing thin-shell instabilities in 3-D simulations. 
Obviously, the exact nature of instabilities will depend on the character of the colliding winds and can be expected to change if different initial parameters are used. 
However, our results make clear that the orbital dynamics of the binary, which cause the split in the nature of the thin-shell instabilities for the LBV+O system, cannot be ignored. 
Future work will consist of performing more parametric studies of the occurrence and dynamical influence of such multiple instabilities. 
   
\begin{acknowledgements}
A.J.v.M.\ acknowledges support from FWO, grant G.0277.08 and K.U.Leuven GOA/09/009. 
Simulations were done at the Flemish High Performance Computer Centre, VIC3 at K.U. Leuven. 
We thank the DEISA Consortium (www.deisa.eu), co-funded through the EU FP6 project RI-031513 and the FP7 project RI-222919, for support within the DEISA Extreme Computing Initiative.\\
We are grateful to our anonymous referee for many helpful comments and suggestions.
\end{acknowledgements}
\vspace{-0.5cm}

\bibliographystyle{aa}
\bibliography{binary_2}

\IfFileExists{\jobname.bbl}{}
 {\typeout{}
  \typeout{******************************************}
  \typeout{** Please run "bibtex \jobname" to obtain}
  \typeout{** the bibliography and then re-run LaTeX}
  \typeout{** twice to fix the references!}
  \typeout{******************************************}
  \typeout{}
 }

\appendix
\section{Turning point of the shell}
\label{sec-turningpoint}
   \begin{figure}
   \centering
   \includegraphics[width=\columnwidth]{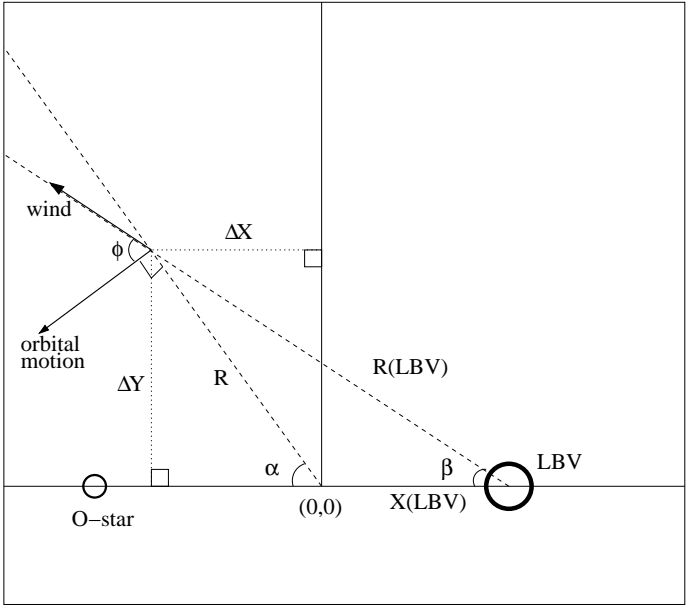}
      \caption{{\bf Schematic overview of the vector components of the slow LBV wind and orbital motion}. The component of the LBV wind in the direction of the orbital motion in a given point ($\Delta\,X,\Delta\,Y$) depends on the  positions of the shell ($\Delta\,X$) and the LBV star ($X(LBV)$) relative to the center of mass (0,0) of the binary.
              }
         \label{fig:binary_sketch_3}
   \end{figure}
For binaries where one of the winds is (relatively) slow compared to the orbital motion, the location of the interaction front between the two winds can be determined more by the relative velocity of this slow wind, than by the point where ram pressure balance exists between the two winds. 
Here we explore the situation where the slower wind is also the stronger in terms of mechanical luminosity. 

\subsection{Mathematical analysis}
The exact point where the trailing end of the shell between the two stars starts to bend away from the ram pressure balance 
is a function of the binary parameters. 
Its location depends on the velocity of the stronger of the two winds, which is pushing it ahead in the orbit, and how it scales against the orbital velocity. 

We will use a 2-D coordinate system with the two stars positioned on the X-axis, rotating counter clockwise, and the center of mass of the binary system in the origin (see Fig.~\ref{fig:binary_sketch_3}). 
Close to the binary stars, the location of the shell can be described as a collection of points ($\Delta\,X, \Delta\,Y$), relative to the center of mass of the binary, which follow the curve along which the ram pressure of the two winds is equal. 
However, as was already mentioned in Sect.~\ref{sec-result}, there is an upper limit to the orbital velocity of the shell: it cannot exceed the velocity of the wind that pushes it. 
(In reality it will always move slower due to the inertia of the shell.) 
Consequently, the shell will start to deviate from the ram pressure balance curve at the point where: $\varv_\angle~\leq~2\pi\,R\Omega $, with $\varv_\angle$ the component of the wind velocity along the direction of orbital motion, $\Omega$ the angular velocity of the binary and $R=\sqrt{\Delta\,X^2 + \Delta\,Y^2}$ the distance from the center of mass to that particular point on the shell. 

This is shown in Fig.~\ref{fig:binary_sketch_3}. 
Here the orbital motion in point $(\Delta\,X, \Delta\,Y)$ and the wind velocity (along the line from the LBV star to the point  $(\Delta\,X, \Delta\,Y$)), make a steep angle, indicating that only a small part of the wind momentum actually pushes the shell ahead in orbit. 
The exact fraction follows from the binary parameters. 
If $\alpha$ is the angle of the wind with the horizontal line between the two stars and $\beta$ is the angle of the position of $(\Delta\,X, \Delta\,Y)$ with the center of mass ($0,0$), the angle of the wind with the orbital motion is given by 
\begin{eqnarray}
\phi~&=&~90^{\rm o} - |{\mathbf \alpha-\beta}| \\
     &=&~90^{\rm o} - \biggl|{\mathbf \sin^{-1}}{\biggl(\frac{\Delta\,Y}{R}\biggr)}-{\mathbf \sin^{-1}}{\biggl(\frac{\Delta\,Y}{R(LBV)}}\biggr)\biggr|,
\end{eqnarray}
where $R(LBV)=\sqrt{\Delta\,Y^2 + (\Delta\,X +X(LBV))^2}$ is the distance from the LBV star to the point on the shell. 
The component of the LBV wind along the direction of orbital motion, $\varv_\angle$ is now given by:
\begin{eqnarray}
\varv_\angle~&=&~ \varv_{\mathrm{LBV}}\cos\biggl(90^{\rm o} -\biggl|\sin^{-1}{\biggl(\frac{\Delta\,Y}{R}\biggr)}-{\mathbf \sin^{-1}}{\biggl(\frac{\Delta\,Y}{R(LBV)}}\biggr)\biggr|\biggr), \\
             &=&~\varv_{\mathrm{LBV}}\sin\biggl(\biggl|{\mathbf \sin^{-1}}{\biggl(\frac{\Delta\,Y}{R}\biggr)}-\sin^{-1}{\biggl(\frac{\Delta\,Y}{R(LBV)}}\biggr)\biggr|\biggr). 
\end{eqnarray}
The condition $\varv_\angle\leq2\pi\,R\Omega$ becomes:
\begin{eqnarray}
2\pi\,R\Omega~&\leq&~\varv_{\mathrm{LBV}}\sin\biggl(\biggl|{\mathbf \sin^{-1}}{\biggl(\frac{\Delta\,Y}{R}\biggr)}-{\mathbf \sin^{-1}}{\biggl(\frac{\Delta\,Y}{R(LBV)}}\biggr)\biggr|\biggr).
\label{eq:v_condition}
\end{eqnarray}
Since $R=\sqrt{\Delta\,X^2 + \Delta\,Y^2}$, this equation can be evaluated for any given position along the X-axis to yield the value $\Delta\,Y$ at which the wind can no longer support the orbital motion. 

The important parameters here are the relative velocities of the rotation and the wind as well as the relative masses of the stars. 
In the most extreme case, where the LBV mass completely dominates the binary, $R(LBV)=R$ so the right hand term of Eq.~\ref{eq:v_condition} goes to zero and the wind would never be able to push the shell ahead in orbit. 
If the LBV actually has a smaller mass than its companion, $R(LBV)\gg\,R$, the region where the wind can push the shell into orbit becomes large.

   \begin{figure}
   \centering
   \includegraphics[width=0.9\columnwidth]{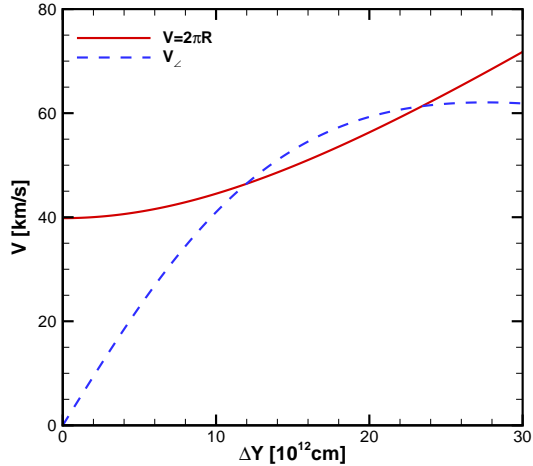}
      \caption{Rotational velocity (continuous) and LBV wind (dashed) along the orbital direction of motion for $\Delta\,X=2\times10^{13}$\,cm in the LBV+O binary. 
The curves cross each other twice, showing the roots of Eq.~\ref{eq:v_condition}. The second root ($\Delta\,Y\simeq2.35\times10^{13}$cm) marks the point where the collision shell starts to deviate from a ram pressure induced bowshock. 
              }
         \label{fig:vcurves}
   \end{figure}

\subsection{Application to a binary simulation}
The condition in Eq.~\ref{eq:v_condition} can be used to determine the point where the shell between the components of a binary system will deviate from the ram pressure balance between the two stellar winds. 

Let us assume, for the moment, that the position of the stars is stationary. 
This is a reasonable assumption for a binary system where the wind velocities are much larger than the orbital velocity. 
Under these conditions the location of the collision front between the stellar winds can be found analytically as a collection of points ($\Delta\,X,\Delta\,Y$) where the ram pressure of the two winds is equal. 
This curve will follow a bow shock that is symmetric around the X-axis. 

We can then use the coordinates of each point on this curve to find the point where this symmetrical bowshock no longer satisfies condition~\ref{eq:v_condition}. At this point the shell will have to deviate significantly from the ram pressure balance, since it has reached its maximum velocity and can not keep up with the rigid rotation. 

As an example we use our simulation of an LBV+O binary. As shown in Figs.~\ref{fig:dens} through  \ref{fig:dens23} the trailing end of the shell runs almost vertical between the two stars. 
Taking for example the X-axis coordinate of the collision point of the two winds directly between the 
two stars ($\Delta\,X\simeq2\times10^{13}$\,cm) and the binary parameters from Table~\ref{tab:bin} we find that the rotational velocity (left side of Eq.~\ref{eq:v_condition}) and the rotational component of the LBV wind velocity (right side of Eq.~\ref{eq:v_condition}) follow the two curves shown in Fig.~\ref{fig:vcurves}. 
Since the equation is second order, there are two roots, one at about $1.20\times10^{13}$\,cm and the second at $2.35\times10^{13}$\,cm.

Therefore, if we would follow the curve of ram pressure balance  we can determine three separate regions. 
Close to the axis connecting the two stars, $\Delta\,Y_{\mathrm shell}$ is lower than the first root of Eq.~\ref{eq:v_condition} for the local value of $\Delta\,X$. 
Here the wind is almost perpendicular to the shell, whereas the orbital motion is almost parallel to the shell. 
Obviously the wind cannot impart orbital motion to the shell, which is effectively standing still, caught between the two winds. 
Since both winds have a momentum component parallel to the shell and counter to the orbital motion (for $\Delta\,Y>0$) the matter is `squeezed out' from between the stars until it reaches the point where $(\Delta\,X,\Delta\,Y)_{\mathrm shell}$ falls between the two roots of the equation.
In this region the shell is caught up by the wind, which can push it into orbit. 
As the matter travels further downstream it reaches the second root, where the LBV wind becomes insufficient and it starts to fall behind the orbital movement of the binary. 

This analysis still ignores part of the orbital motion in that it assumes that the wind impacting on the shell at a given time $t$ was also launched from the star at that same time $t$. 
In reality the wind was emitted earlier, which means that the star was at a different position. How important this is depends on the total displacement of the star between the time the wind is launched and the time it hits the shell.

\end{document}